\documentclass[titlepage]{article}%
\usepackage{amsmath}
\usepackage{cite}%
\usepackage{amsfonts}%
\usepackage{amssymb}%
\usepackage{graphicx}
\newtheorem{theorem}{Theorem}

\newtheorem{lemma}[theorem]{Lemma}

\newenvironment{proof}[1][Proof]{\noindent\textbf{#1.} }{\ \rule{0.5em}{0.5em}}
\begin{document}

\title{Second look at the spread of epidemics on networks}
\author{Eben Kenah$^{\ast}$, James M. Robins\\Departments of Epidemiology and Biostatistics\\Harvard School of Public Health\\677 Huntington Avenue, Boston, MA 02115\\$^{\ast}$Corresponding author: ekenah@hsph.harvard.edu}
\date{Physical Review E \textbf{76,} 036113 (2007)}
\maketitle

\begin{abstract}
In an important paper, M.E.J. Newman claimed that a general network-based
stochastic Susceptible-Infectious-Removed (SIR) epidemic model is isomorphic
to a bond percolation model, where the bonds are the edges of the contact
network and the bond occupation probability is equal to the marginal
probability of transmission from an infected node to a susceptible neighbor.
\ In this paper, we show that this isomorphism is incorrect and define a
semi-directed random network we call the \textit{epidemic} \textit{percolation
network} that is exactly isomorphic to the SIR epidemic model in any finite
population. \ In the limit of a large population, (i) the distribution of
(self-limited) outbreak sizes is identical to the size distribution of (small)
out-components, (ii) the epidemic threshold corresponds to the phase
transition where a giant strongly-connected component appears, (iii) the
probability of a large epidemic is equal to the probability that an initial
infection occurs in the giant in-component, and (iv) the relative final size
of an epidemic is equal to the proportion of the network contained in the
giant out-component. \ For the SIR model considered by Newman, we show that
the epidemic percolation network predicts the same mean outbreak size below
the epidemic threshold, the same epidemic threshold, and the same final size
of an epidemic as the bond percolation model. \ However, the bond percolation
model fails to predict the correct outbreak size distribution and probability
of an epidemic when there is a nondegenerate infectious period distribution.
\ We confirm our findings by comparing predictions from percolation networks
and bond percolation models to the results of simulations. \ In an appendix,
we show that an isomorphism to an epidemic percolation network can be defined
for any time-homogeneous stochastic SIR model.

\end{abstract}

\section{Introduction}

In an important paper, M. E. J. Newman studied a network-based
Susceptible-Infectious-Removed (SIR) epidemic model in which infection is
transmitted through a network of contacts between individuals \cite{Newman1}.
\ The contact network itself is a random undirected network with an arbitrary
degree distribution of the form studied by Newman, Strogatz, and Watts
\cite{NSW}. \ Given the degree distribution, these networks are maximally
random, so they have no small loops and no degree correlations in the limit of
a large population \cite{NSW,Boguna,Meyers}. \ 

In the stochastic SIR model considered by Newman, the probability that an
infected node $i$ makes infectious contact with a neighbor $j$ is given by
$T_{ij}=1-\exp(-\beta_{ij}\tau_{i})$, where $\beta_{ij}$ is the rate of
infectious contact from $i$ to $j$ and $\tau_{i}$ is the time that $i$ remains
infectious. \ (We use \textit{infectious contact} to mean a contact that
results in infection if and only if the recipient is susceptible.) \ The
infectious period $\tau_{i}$ is a random variable with the cumulative
distribution function (cdf) $F(\tau)$, and the infectious contact rate
$\beta_{ij}$ is a random variable with the cdf $F(\beta)$. \ The infectious
periods for all individuals are independent and identically distributed (iid),
and the infectious contact rates for all ordered pairs of individuals are iid. \ 

Under these assumptions, Newman claimed that the spread of disease on the
contact network is exactly isomorphic to a bond percolation model on the
contact network with bond occupation probability equal to the \textit{a
priori} probability of disease transmission between any two connected nodes in
the contact network \cite{Newman1}. \ This probability is called the
\textit{transmissibility} and denoted by $T$:%
\begin{equation}
T=\left\langle T_{ij}\right\rangle =\int_{0}^{\infty}\int_{0}^{\infty
}(1-e^{-\beta_{ij}\tau_{i}})dF(\beta_{ij})dF(\tau_{i}).\label{T}%
\end{equation}
Newman used this bond percolation model to derive the distribution of finite
outbreak sizes, the critical transmissibility $T_{c}$ that defines the
epidemic (i.e., percolation) threshold, and the probability and relative final
size of an epidemic (i.e., an outbreak that never goes extinct). \ \ 

As a counterexample, consider a contact network where each subject has exactly
two contacts. \ Assume that (i) $\tau_{i}=\tau_{0}>0$ with probability $p$ and
$\tau_{i}=0$ with probability $1-p$ and (ii) $\beta_{ij}=\beta_{0}>0$ with
probability one for all $ij$. \ Under the SIR model, the probability that the
infection of a randomly chosen node results in an outbreak of size one is
$p_{1}=1-p+pe^{-2\beta_{0}\tau_{0}}$, which is the sum of the probability
$1-p$ that $\tau=0$ and the probability $pe^{-2\beta_{0}\tau_{0}}$ that
$\tau=\tau_{0}$ and disease is not transmitted to either contact. \ Under the
bond percolation model, the probability of a cluster of size one is
$p_{1}^{bond}=(1-p+pe^{-\beta_{0}\tau_{0}})^{2}$, corresponding to the
probability that neither of the bonds incident to the node are occupied.
\ Since
\[
p_{1}-p_{1}^{bond}=p(1-p)(1-e^{-\beta_{0}\tau_{0}})^{2},
\]
the bond percolation model correctly predicts the probability of an outbreak
of size one only if $p=0$ or $p=1$. \ When the infectious period is not
constant, it underestimates this probability. \ The supremum of the error is
$0.25$, which occurs when $p=0.5$ and $\tau_{0}\rightarrow\infty$. \ In this
limit, the SIR model corresponds to a site percolation model rather than a
bond percolation model. \ 

When the distribution of infectious periods is nondegenerate, there is no bond
occupation probability that will make the bond percolation model isomorphic to
the SIR model. \ To see why, suppose node $i$ has infectious period $\tau_{i}$
and degree $n_{i}$ in the contact network. \ In the epidemic model, the
conditional probability that $i$ transmits infection to a neighbor $j$ in the
contact network given $\tau_{i}$ is
\begin{equation}
T_{\tau_{i}}=\int_{0}^{\infty}(1-e^{-\beta_{ij}\tau_{i}})dF(\beta
_{ij}).\label{Ti}%
\end{equation}
Since the contact rate pairs for all $n_{i}$ edges incident to $i$ are iid,
the transmission events across these edges are (conditionally) independent
Bernoulli($T_{\tau_{i}}$) random variables; but the transmission probabilities
are strictly increasing in $\tau_{i}$, so the transmission events are
(marginally) dependent unless $\tau_{i}=\tau_{0}$ with probability one for
some fixed $\tau_{0}$. \ In contrast, the bond percolation model\ treats the
infections generated by node $i$ as $n_{i}$ (marginally) independent
Bernoulli($T$) random variables regardless of the distribution of $\tau_{i}$.
\ Neither counterexample assumes anything about the global properties of the
contact network, so Newman's claim cannot be justified as an approximation in
the limit of a large network with no small loops. \ 

In Section 2, we define a semi-directed random network called the
\textit{epidemic percolation network} and show how it can be used to predict
the outbreak size distribution, the epidemic threshold, and the probability
and final size of an epidemic in the limit of a large population for any
time-homogeneous SIR model. \ In Section 3, we show that the network-based
stochastic SIR model from \cite{Newman1} can be analyzed correctly using a
semi-directed random network of the type studied by Bogu\~{n}\'{a} and Serrano
\cite{Boguna}. \ In Section 4, we show that it predicts the same epidemic
threshold, mean outbreak size below the epidemic threshold, and relative final
size of an epidemic as the bond percolation model. \ In Section 5, we show
that the bond percolation model fails to predict the distribution of outbreak
sizes and the probability of an epidemic when the distribution of infectious
periods is nondegenerate. \ In Section 6, we compare predictions made by
epidemic percolation networks and bond percolation models to the results of
simulations. \ In an appendix, we define epidemic percolation networks for a
very general time-homogeneous stochastic SIR model and show that their
out-components are isomorphic to the distribution of possible outcomes of the
SIR model for any given set of imported infections.

\section{Epidemic percolation networks}

Consider a node $i$ with degree $n_{i}$ in the contact network and infectious
period $\tau_{i}$. \ In the SIR model defined above, the number of people who
will transmit infection to $i$ if they become infectious has a binomial($n_{i}%
,T$) distribution regardless of $\tau_{i}$. \ If $i$ is infected along one of
the $n_{i}$ edges, then the number of people to whom $i $ will transmit
infection has a binomial($n_{i}-1,T_{\tau_{i}}$) distribution. \ In order to
produce the correct joint distribution of the number of people who will
transmit infection to $i$ and the number of people to whom $i$ will transmit
infection, we represent the former by directed edges that terminate at $i$ and
the latter by directed edges that originate at $i$. \ Since there can be at
most one transmission of infection between any two persons, we replace pairs
of directed edges between two nodes with a single undirected edge. \ 

Starting from the contact network, a single realization of the
\textit{epidemic percolation network} can be generated as follows:

\begin{enumerate}
\item Choose a recovery period $\tau_{i}$ for every node $i$ in the network
and choose a contact rate $\beta_{ij}$ for every ordered pair of connected
nodes $i$ and $j$ in the contact network.

\item For each pair of connected nodes $i$ and $j$ in the contact network,
convert the undirected edge between them to a directed edge from $i$ to $j$
with probability $(1-e^{-\beta_{ij}\tau_{i}})e^{-\beta_{ji}\tau_{j}}$, to a
directed edge from $j$ to $i$ with probability $e^{-\beta_{ij}\tau_{i}%
}(1-e^{-\beta_{ji}\tau_{j}})$, and erase the edge completely with probability
$e^{-\beta_{ij}\tau_{i}-\beta_{ji}\tau_{j}}$. \ The edge remains undirected
with probability $(1-e^{-\beta_{ij}\tau_{i}})(1-e^{-\beta_{ji}\tau_{j}})$. \ 
\end{enumerate}

The epidemic percolation network is a semi-directed random network that
represents a single realization of the infectious contact process for each
connected pair of nodes, so $4^{m}$ possible percolation networks exist for a
contact network with $m$ edges. \ The probability of each possible network is
determined by the underlying SIR model. \ The epidemic percolation network is
very similar to the locally dependent random graph defined by Kuulasmaa
\cite{Kuulasmaa1} for an epidemic on a $d$-dimensional lattice. \ There are
two important differences: First, the underlying structure of the contact
network is not assumed to be a lattice. \ Second, we replace pairs of
(occupied) directed edges between two nodes with a single undirected edge so
that its component structure can be analyzed using a generating function
formalism. \ 

In the Appendix, we prove that the size distribution of outbreaks starting
from any node in a time-homogeneous stochastic SIR model is identical to the
distribution of its out-component sizes in the corresponding probability space
of percolation networks. \ Since this result applies to any time-homogeneous
SIR\ model, it can be used to analyze network-based models, fully-mixed models
(see \cite{Kenah}), and models with multiple levels of mixing. \ 

\subsection{Components of semi-directed networks}

In this section, we briefly review the structure of directed and semi-directed
networks as discussed in \cite{Boguna,Meyers,Broder,Dorogovtsev}. \ In the
next section, we relate this to the possible outcomes of an SIR\ model. \ 

The \textit{indegree} and \textit{outdegree} of node $i$ are the number of
incoming and outgoing directed edges incident to $i$. \ Since each directed
edge is an outgoing edge for one node and an incoming edge for another node,
the mean indegree and outdegree are equal. \ The \textit{undirected degree} of
node $i$ is the number of undirected edges incident to $i$. \ The
\textit{size} of a component is the number of nodes it contains and its
\textit{relative size} is its size divided by the total size of the network. \ 

The \textit{out-component} of node $i$ includes $i$ and all nodes that can be
reached from $i$ by following a series of edges in the proper direction
(undirected edges are bidirectional). \ The \textit{in-component} of node $i$
includes $i$ and all nodes from which $i$ can be reached by following a series
of edges in the proper direction. \ By definition, node $i$ is in the
in-component of node $j$ if and only if $j$ is in the out-component of $i$.
\ Therefore, the mean in- and out-component sizes in any (semi-)directed
network are equal. \ 

The \textit{strongly-connected component} of a node $i$ is the intersection of
its in- and out-components; it is the set of all nodes that can be reached
from node $i$ and from which node $i$ can be reached. \ All nodes in a
strongly-connected component have the same in-component and the same
out-component. \ The \textit{weakly-connected component} of node $i$ is the
set of nodes that are connected to $i$ when the direction of the edges is
ignored. \ 

\begin{figure}
[ptb]
\begin{center}
\includegraphics[
height=4.1684in,
width=5.5486in
]%
{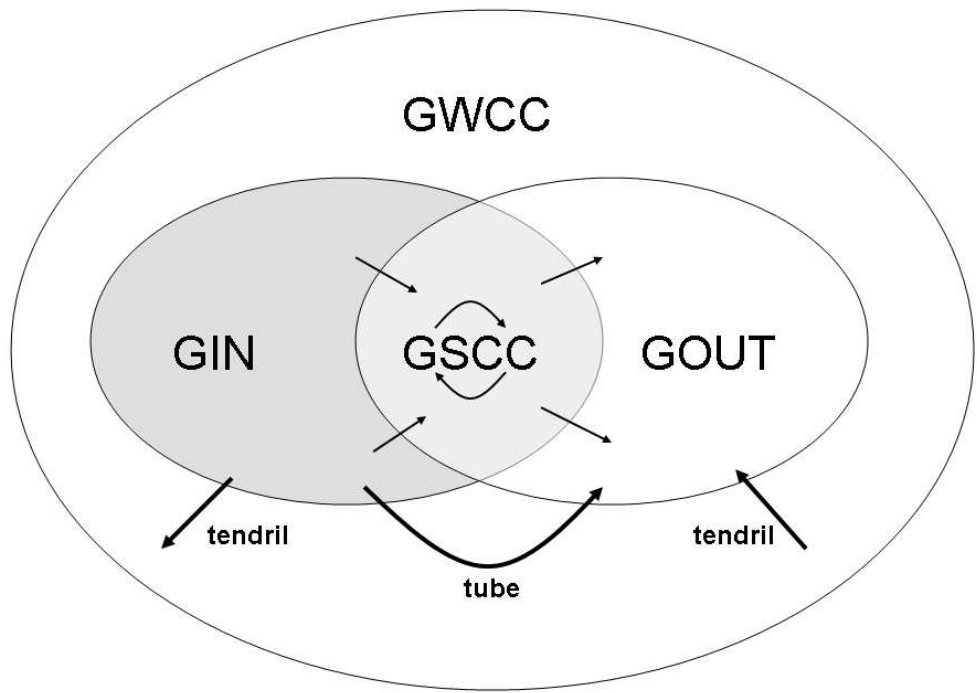}%
\caption{Schematic diagram of the giant components, tendrils, and tubes of a
supercritical semi-directed network. \ Adapted from Broder \textit{et al.}
\cite{Broder} and Dorogovtsev \textit{et al.} \cite{Dorogovtsev}. }%
\label{bowtie}%
\end{center}
\end{figure}

For giant components, we use the definitions given in
\cite{Dorogovtsev,Schwartz}. \ Giant components have asymptotically positive
relative size in the limit of a large population. \ All other components are
\textquotedblleft small" in the sense that they have asymptotically zero
relative size. \ There are two phase transitions in a semi-directed network:
One where a unique giant weakly-connected component (GWCC) emerges and another
where unique giant in-, out-, and strongly-connected components (GIN, GOUT,
and GSCC) emerge. \ The GWCC contains the other three giant components. \ The
GSCC is the intersection of the GIN and the GOUT, which are the common in- and
out-components of nodes in the GSCC. \ \textit{Tendrils} are components in the
GWCC that are outside the GIN and the GOUT. \ \textit{Tubes} are directed
paths from the GIN to the GOUT that do not intersect the GSCC. \ All tendrils
and tubes are small components. \ A schematic representation of these
components is shown in Figure (\ref{bowtie}).

\subsection{Epidemic percolation networks and epidemics}

An \textit{outbreak} begins when one or more nodes are infected from outside
the population. \ These are called \textit{imported infections}. \ The
\textit{final size} of an outbreak is the number of nodes that are infected
before the end of transmission, and its \textit{relative final size} is its
final size divided by the total size of the network. \ In the epidemic
percolation network, the nodes infected in the outbreak can be identified with
the nodes in the out-components of the imported infections. \ This
identification is made mathematically precise in the Appendix. \ 

Informally, we define a \textit{self-limited outbreak} to be an outbreak whose
relative final size approaches zero in the limit of a large population and an
\textit{epidemic} to be an outbreak whose relative final size is positive in
the limit of a large population. \ There is a critical transmissibility
$T_{c}$ that defines the \textit{epidemic threshold:} The probability of an
epidemic is zero when $T\leq T_{c}$, and the probability and final size of an
epidemic are positive when $T>T_{c}$ \cite{Newman1,Andersson,Diekmann,Sander}. \ 

If all out-components in the epidemic percolation network are small, then only
self-limited outbreaks are possible. \ If the percolation network contains a
GSCC, then any infection in the GIN will lead to the infection of the entire
GOUT. \ Therefore, the epidemic threshold corresponds to the emergence of the
GSCC in the percolation network. \ For any finite set of imported infections,
the probability of an epidemic is equal to the probability that at least one
imported infection occurs in the GIN. \ The relative final size of an epidemic
is equal to the proportion of the network contained in the GOUT. \ Although
some nodes outside the GOUT may be infected (e.g. nodes in tendrils and
tubes), they constitute a finite number of small components whose total
relative size is asymptotically zero.

\section{Analysis of the SIR model}

To analyze the SIR model from \cite{Newman1}, we first calculate the
probability generating function (pgf) of the degree distribution of the
corresponding epidemic percolation network. \ Then we use methods developed by
Bogu\~{n}\'{a} and Serrano \cite{Boguna}\ and Meyers \textit{et al.}
\cite{Meyers} to calculate the in- and out-component size distributions and
the relative sizes of the GIN, GOUT, and GSCC.

\subsection{Degree distribution}

If $p_{n}$ is the probability that a node has degree $n$ in the contact
network, then
\[
\mathcal{G}(z)=\sum_{n=1}^{\infty}p_{n}z^{n}%
\]
is the probability generating function (pgf) for the degree distribution of
the contact network. \ If $p_{jkm}$ is the probability that a node in the
epidemic percolation network has $j$ incoming edges, $k$ outgoing edges, and
$m$ undirected edges, then
\[
G(x,y,u)=\sum_{j=0}^{\infty}\sum_{k=0}^{\infty}\sum_{m=0}^{\infty}p_{jkm}%
x^{j}y^{k}u^{m}%
\]
is the pgf for the degree distribution of the percolation network. \ Suppose
nodes $i$ and $j$ are connected in the contact network with contact rates
$(\beta_{ij},\beta_{ji})$ and infectious periods $\tau_{i}$ and $\tau_{j}$.
\ Let $g(x,y,u|\beta_{ij},\beta_{ji},\tau_{i},\tau_{j})$ be the conditional
pgf for the number of incoming, outgoing, and undirected edges incident to $i$
that appear between $i$ and $j$ in the percolation network. \ Then
\begin{align*}
g(x,y,u|\beta_{ij},\beta_{ji},\tau_{i},\tau_{j})  & =e^{-\beta_{ij}\tau
_{i}-\beta_{ji}\tau_{j}}+e^{-\beta_{ij}\tau_{i}}(1-e^{-\beta_{ji}\tau_{j}%
})x\\
& +(1-e^{-\beta_{ij}\tau_{i}})e^{-\beta_{ji}\tau_{j}}y+(1-e^{-\beta_{ij}%
\tau_{i}})(1-e^{-\beta_{ji}\tau_{j}})u.
\end{align*}
Given $\tau_{i}$, the conditional pgf for the number of incoming, outgoing,
and undirected edges incident to $i$ that appear in the percolation network
between $i$ and any neighbor of $i$ in the contact network is
\begin{align}
g(x,y,u|\tau_{i})  & =\int_{0}^{\infty}\int_{0}^{\infty}\int_{0}^{\infty
}g(x,y,u|\beta_{ij},\beta_{ji},\tau_{i},\tau_{j})dF(\beta_{ij})dF(\beta
_{ji})dF(\tau_{j})\nonumber\\
& =(1-T_{\tau_{i}})(1-T)+(1-T_{\tau_{i}})Tx+T_{\tau_{i}}(1-T)y+T_{\tau_{i}%
}Tu.\label{gxyu|tau}%
\end{align}
The pgf for the degree distribution of a node with infectious period $\tau
_{i}$ is
\begin{equation}
G(x,y,u|\tau_{i})=\sum_{n=0}^{\infty}p_{n}(g(x,y,u|\tau_{i}))^{n}%
=\mathcal{G}(g(x,y,u|\tau_{i})).\label{Gxyu|tau}%
\end{equation}
Finally, the pgf for the degree distribution of the epidemic percolation
network is
\begin{equation}
G(x,y,u)=\int_{0}^{\infty}G(x,y,u|\tau_{i})dF(\tau_{i}).\label{Gxyu}%
\end{equation}

If $a$, $b$, and $c$ are nonnegative integers, let $G^{(a,b,c)}(x,y,u)$ be the
derivative obtained after differentiating $a$ times with respect to $x$, $b$
times with respect to $y$, and $c$ times with respect to $u$. \ Then the mean
indegree and outdegree of the percolation network are
\[
\left\langle k_{d}\right\rangle =G^{(1,0,0)}(1,1,1)=G^{(0,1,0)}%
(1,1,1)=T(1-T)\mathcal{G}^{\prime}(1),
\]
and the mean undirected degree is
\[
\left\langle k_{u}\right\rangle =G^{(0,0,1)}(1,1,1)=T^{2}\mathcal{G}^{\prime
}(1).
\]
\ 

\subsection{Generating functions}

When the contact network underlying an SIR epidemic model is an undirected
random network with an arbitrary degree distribution, the pgf of its degree
distribution can be used to calculate the distribution of small component
sizes, the percolation threshold, and the relative sizes of the GIN, GOUT, and
GSCC using methods developed by Bogu\~{n}\'{a} and Serrano \cite{Boguna}\ and
Meyers \textit{et al.} \cite{Meyers}. \ These methods generalize earlier
methods for undirected and purely directed networks
\cite{NSW,Newman1,Albert,Newman2,Newman3,NBW}. \ In this section, we review
these results and introduce notation that will be used in the rest of the
paper. \ We discuss the case of networks with no two-point degree
correlations, which is sufficient to analyze the SIR model from \cite{Newman1}.

Let $G_{f}(x,y,u)$ be the pgf for the degree distribution of a node reached by
going forward along a directed edge, excluding the edge used to reach the
node. \ Since the probability of reaching any node by following a directed
edge is proportional to its indegree,
\begin{equation}
G_{f}(x,y,u)=\frac{1}{\left\langle k_{d}\right\rangle }\sum_{j,k,m}%
jp_{jkm}x^{j-1}y^{k}u^{m}=\frac{1}{\left\langle k_{d}\right\rangle
}G^{(1,0,0)}(x,y,u).\label{Gf*}%
\end{equation}
Similarly, the pgf for the degree distribution of a node reached by going in
reverse along a directed edge (excluding the edge used to reach the node) is
\begin{equation}
G_{r}(x,y,u)=\frac{1}{\left\langle k_{d}\right\rangle }G^{(0,1,0)}%
(x,y,u),\label{Gr*}%
\end{equation}
and the pgf for the degree distribution of a node reached by going to the end
of an undirected edge (excluding the edge used to reach the node) is
\begin{equation}
G_{u}(x,y,u)=\frac{1}{\left\langle k_{u}\right\rangle }G^{(0,0,1)}%
(x,y,u).\label{Gu*}%
\end{equation}
\ 

\subsubsection{Out-components}

Let $H_{f}^{out}(z)$ be the pgf for the size of the out-component at the end
of a directed edge and $H_{u}^{out}(z)$ be the pgf for the size of the
out-component at the \textquotedblleft end" of an undirected edge. \ Then, in
the limit of a large population,
\begin{subequations}
\begin{align}
H_{f}^{out}(z)  & =zG_{f}(1,H_{f}^{out}(z),H_{u}^{out}(z)),\label{Hf,out}\\
H_{u}^{out}(z)  & =zG_{u}(1,H_{f}^{out}(z),H_{u}^{out}(z)).\label{Hu,out}%
\end{align}
The pgf for the out-component size of a randomly chosen node is
\end{subequations}
\begin{equation}
H^{out}(z)=zG(1,H_{f}^{out}(z),H_{u}^{out}(z)).\label{Hout}%
\end{equation}
The probability that a node has a finite out-component in the limit of a large
population is $H^{out}(1)$, so the probability that a randomly chosen node is
in the GIN is $1-H^{out}(1)$. \ 

The coefficients on $z^{0}$ in $H_{f}^{out}(z)$ and $H_{u}^{out}(z)$ are
$G_{f}(1,0,0)$ and $G_{u}(1,0,0)$ respectively. \ Therefore, power series for
$H_{f}^{out}(z)$ and $H_{u}^{out}(z)$ can be computed to any desired order by
iterating equations (\ref{Hf,out}) and (\ref{Hu,out}). \ A power series for
$H^{out}(z)$ can then be obtained using equation (\ref{Hout}). \ For any
$z\in\lbrack0,1]$, $H_{f}^{out}(z)$ and $H_{u}^{out}(z)$ can be calculated
with arbitrary precision by iterating equations (\ref{Hf,out}) and
(\ref{Hu,out}) starting from initial values $y_{0},u_{0}\in\lbrack0,1)$.
\ Estimates of $H_{f}^{out}(z)$ and $H_{u}^{out}(z)$ can be used to estimate
$H^{out}(z)$ with arbitrary precision. \ 

The expected size of the out-component of a randomly chosen node below the
epidemic threshold is $H^{out\prime}(1)$. \ Taking derivatives in (\ref{Hout})
yields
\begin{equation}
H^{out\prime}(1)=1+\left\langle k_{d}\right\rangle H_{f}^{out\prime
}(1)+\left\langle k_{u}\right\rangle H_{u}^{out\prime}(1).\label{Hout'(1)}%
\end{equation}
Taking derivatives in equations (\ref{Hf,out}) and (\ref{Hu,out}) and using
the fact that $H_{f}^{out}(1)=H_{u}^{out}(1)=1$ below the epidemic threshold
yields a set of linear equations for $H_{f}^{out\prime}(1)$ and $H_{u}%
^{out\prime}(1)$. \ These can be solved to yield
\begin{equation}
H_{f}^{out\prime}(1)=\frac{1+G_{f}^{(0,0,1)}-G_{u}^{(0,0,1)}}{(1-G_{f}%
^{(0,1,0)})(1-G_{u}^{(0,0,1)})-G_{f}^{(0,0,1)}G_{u}^{(0,1,0)}}%
\label{Hfout'(1)}%
\end{equation}
and
\begin{equation}
H_{u}^{out\prime}(1)=\frac{1-G_{f}^{(0,1,0)}+G_{u}^{(0,1,0)}}{(1-G_{f}%
^{(0,1,0)})(1-G_{u}^{(0,0,1)})-G_{f}^{(0,0,1)}G_{u}^{(0,1,0)}}%
,\label{Huout'(1)}%
\end{equation}
where the argument of all derivatives is $(1,1,1)$. \ 

\subsubsection{In-components}

The in-component size distribution of a semi-directed network can be derived
using the same logic used to find the out-component size distribution, except
that we consider going backwards along directed edges. \ Let $H_{r}^{in}(z)$
be the pgf for the size of the in-component at the beginning of a directed
edge, $H_{u}^{in}(z)$ be the pgf for the size of the in-component at the
\textquotedblleft beginning" of an undirected edge, and $H^{in}(z)$ be the pgf
for the in-component size of a randomly chosen node. \ Then, in the limit of a
large population,
\begin{subequations}
\label{Hur,in}%
\begin{align}
H_{r}^{in}(z)  & =zG_{r}(H_{r}^{in}(z),1,H_{u}^{in}(z)),\label{Hr,in}\\
H_{u}^{in}(z)  & =zG_{u}(H_{r}^{in}(z),1,H_{u}^{in}(z)),\label{Hu,in}\\
H^{in}(z)  & =zG(H_{r}^{in}(z),1,H_{u}^{in}(z)).\label{Hin}%
\end{align}
The probability that a node has a finite in-component is $H^{in}(1)$, so the
probability that a randomly chosen node is in the GOUT is $1-H^{in}(1)$. \ The
expected size of the in-component of a randomly chosen node is $H^{in\prime
}(1)$. \ Power series and numerical estimates for $H_{r}^{in}(z)$, $H_{u}%
^{in}(z)$, and $H^{in}(z)$ can be obtained by iterating these equations.

The expected size of the out-component of a randomly chosen node below the
epidemic threshold is $H^{out\prime}(1)$. \ Taking derivatives in equation
(\ref{Hin}) yields
\end{subequations}
\begin{equation}
H^{in\prime}(1)=1+\left\langle k_{d}\right\rangle H_{r}^{in\prime
}(1)+\left\langle k_{u}\right\rangle H_{u}^{in\prime}(1).\label{Hin'(1)}%
\end{equation}
Taking derivatives in equations (\ref{Hr,in}) and (\ref{Hu,in}) and using the
fact that $H_{r}^{in}(1)=H_{u}^{in}(1)=1$ in a subcritical network yields
\begin{equation}
H_{r}^{in\prime}(1)=\frac{1+G_{r}^{(0,0,1)}-G_{u}^{(0,0,1)}}{(1-G_{r}%
^{(1,0,0)})(1-G_{u}^{(0,0,1)})-G_{r}^{(0,0,1)}G_{u}^{(1,0,0)}}\label{Hrin'(1)}%
\end{equation}
and
\begin{equation}
H_{u}^{in\prime}(1)=\frac{1-G_{r}^{(1,0,0)}+G_{u}^{(1,0,0)}}{(1-G_{r}%
^{(1,0,0)})(1-G_{u}^{(0,0,1)})-G_{r}^{(0,0,1)}G_{u}^{(1,0,0)}}%
,\label{Huin'(1)}%
\end{equation}
where the argument of all derivatives is $(1,1,1)$. \ 

\subsubsection{Epidemic threshold}

The epidemic threshold occurs when the expected size of the in- and
out-components in the network becomes infinite. \ This occurs when the
denominators in equations (\ref{Hfout'(1)}) and (\ref{Huout'(1)}) and
equations (\ref{Hrin'(1)}) and (\ref{Huin'(1)}) approach zero. \ From the
definitions of $G_{f}(x,y,u)$, $G_{r}(x,y,u)$ and $G_{u}(x,y,u)$, both
conditions are equivalent to
\[
(1-\frac{1}{\left\langle k_{d}\right\rangle }G^{(1,1,0)})(1-\frac
{1}{\left\langle k_{u}\right\rangle }G^{(0,0,2)})-\frac{1}{\left\langle
k_{d}\right\rangle \left\langle k_{u}\right\rangle }G^{(1,0,1)}G^{(0,1,1)}=0.
\]
Therefore, there is a single epidemic threshold where the GSCC, the GIN, and
the GOUT appear simultaneously in both purely directed networks
\cite{NSW,Newman1,Albert,Newman2,Newman3,NBW} and semi-directed networks
\cite{Boguna,Meyers}.

\subsubsection{Giant strongly-connected component}

A node is in the GSCC if its in- and out-components are both infinite. \ A
randomly chosen node has a finite in-component with probability $G(H_{r}%
^{in}(1),1,H_{u}^{in}(1))$ and a finite out-component with probability
$G(1,H_{f}^{out}(1),H_{u}^{out}(1))$. \ The probability that a node reached by
following an undirected edge has finite in- and out-components is the solution
to the equation
\[
v=G_{u}(H_{r}^{in}(1),H_{f}^{out}(1),v),
\]
and the probability that a randomly chosen node has finite in- and
out-components is $G(H_{r}^{in}(1),H_{f}^{out}(1),v)$ \cite{Boguna}. \ Thus,
the relative size of the GSCC is
\[
1-G(H_{r}^{in}(1),1,H_{u}^{in}(1))-G(1,H_{f}^{out}(1),H_{u}^{out}%
(1))+G(H_{r}^{in}(1),H_{f}^{out}(1),v).
\]

\section{In-components}

In this section, we prove that the in-component size distribution of the
epidemic percolation network for the SIR model from \cite{Newman1} is
identical to the component size distribution of the bond percolation model
with bond occupation probability $T$. \ The probability generating function
for the total number of incoming and undirected edges incident to any node $i$
is
\[
G(x,1,x|\tau_{i})=\mathcal{G}(g(x,1,x|\tau_{i}))=\mathcal{G}(1-T+Tx),
\]
which is independent of $\tau_{i}$. \ If node $i$ has degree $n_{i}$ in the
contact network, then the number of nodes we can reach by going in reverse
along a directed edge or an undirected edge has a binomial$(n_{i},T)$
distribution regardless of $\tau_{i}$. \ If we reach node $i$ by going
backwards along edges, the number of nodes we can reach from $i$ by continuing
to go backwards (excluding the node from which we arrived) has a
binomial$(n_{i}-1,T)$ distribution. \ Therefore, the in-component of any node
in the percolation network is exactly like a component of a bond percolation
model with occupation probability $T$. \ This argument was used to justify the
mapping from an epidemic model to a bond percolation model in \cite{Newman1},
but it does not apply to the out-components of the epidemic percolation
network. \ 

Methods of calculating the component size distribution of an undirected random
network with an arbitrary degree distribution using the pgf of its degree
distribution were developed by Newman \textit{et al}. \cite{NSW,
Albert,Newman2,Newman3,NBW}. \ These methods were used to analyze the bond
percolation model of disease transmission \cite{Newman1}, obtaining results
similar to those obtained by Andersson \cite{Andersson2} for the epidemic
threshold and the final size of an epidemic. \ In this paragraph, we review
these results and introduce notation that will be used in this section. \ Let
$\mathcal{G}(u)$ be the pgf for the degree distribution of the contact
network. \ Then the pgf for the degree of a node reached by following an edge
(excluding the edge used to reach that node) is $\mathcal{G}_{1}%
(u)=\left\langle n\right\rangle ^{-1}\mathcal{G}^{\prime}(u)$, where
$\left\langle n\right\rangle =\mathcal{G}^{\prime}(1)$ is the mean degree of
the contact network. \ With bond occupation probability $T$, the number of
occupied edges adjacent to a randomly chosen node has the pgf $\mathcal{G}%
(1-T+Tu)$ and the number of occupied edges from which infection can leave a
node that has been infected along an edge has the pgf $\mathcal{G}%
_{1}(1-T+Tu)$. \ The pgf for the size of the component at the end of an edge
is
\begin{equation}
H_{1}(z)=z\mathcal{G}_{1}(1-T+TH_{1}(z))\label{H1}%
\end{equation}
and the pgf for the size of the component of a randomly chosen node is
\begin{equation}
H_{0}(z)=z\mathcal{G}(1-T+TH_{1}(z)).\label{H0}%
\end{equation}
The proportion of the network contained in the giant component is $1-H_{0}(1)
$, and the mean size of components below the percolation threshold is
$H_{0}^{\prime}(1)$. \ $H_{0}(z)$ and $H_{1}(z)$ can be expanded as power
series to any desired degree by iterating equations (\ref{H1}) and
(\ref{H0}),\ and their value for any fixed $z\in\lbrack0,1]$ can be found by
iteration from an initial value $z_{0}\in\lbrack0,1)$. \ 

We can now prove that the distribution of component sizes in the bond
percolation model is identical to the distribution of in-component sizes in
the epidemic percolation network.

\begin{lemma}
$G_{r}(x,y,u)=G_{u}(x,y,u)$ for all $x,y,u.$
\end{lemma}

\begin{proof}
From equation (\ref{Gr*}),
\begin{align*}
G_{r}(x,y,u)  & =\frac{1}{T(1-T)\mathcal{G}^{\prime}(1)}G^{(0,1,0)}(x,y,u)\\
& =\frac{1}{T\mathcal{G}^{\prime}(1)}\int_{0}^{\infty}\mathcal{G}^{\prime
}(g(x,y,u|\tau_{i}))T_{\tau_{i}}dF(\tau_{i}).
\end{align*}
From equation (\ref{Gu*}),%
\begin{align*}
G_{u}(x,y,u)  & =\frac{1}{T^{2}\mathcal{G}^{\prime}(1)}G^{(0,0,1)}(x,y,u)\\
& =\frac{1}{T\mathcal{G}^{\prime}(1)}\int_{0}^{\infty}\mathcal{G}^{\prime
}(g(x,y,u|\tau_{i}))T_{\tau_{i}}dF(\tau_{i})\text{.}%
\end{align*}
Thus, the degree distribution of a node reached by going backwards along an
edge is independent of whether it was a directed or undirected edge. \ 
\end{proof}

\begin{lemma}
$H_{r}^{in}(z)=H_{u}^{in}(z)=H_{1}(z)$ for all $z$. \ 
\end{lemma}

\begin{proof}
From equations (\ref{Hr,in}) and (\ref{Hu,in}),
\begin{align*}
H_{r}^{in}(z)  & =zG_{r}(H_{r}^{in}(z),1,H_{u}^{in}(z))\\
& =zG_{u}(H_{r}^{in}(z),1,H_{u}^{in}(z))=H_{u}^{in}(z)\text{.}%
\end{align*}
Let $H_{\ast}^{in}(z)=H_{u}^{in}(z)=H_{r}^{in}(z)$. \ Since $g(x,1,x|\tau
_{i})=1-T+Tx$ for all $\tau_{i}$,
\begin{align*}
H_{\ast}^{in}(z)  & =\frac{z}{T\mathcal{G}^{\prime}(1)}\int_{0}^{\infty
}\mathcal{G}^{\prime}(1-T+TH_{\ast}^{in}(z))T_{\tau_{i}}dF(\tau_{i})\\
& =\frac{z}{\mathcal{G}^{\prime}(1)}\mathcal{G}^{\prime}(1-T+TH_{\ast}%
^{in}(z))\text{.}%
\end{align*}
From equation (\ref{H1}), we have
\[
H_{1}(z)=\frac{z}{\mathcal{G}^{\prime}(1)}\mathcal{G}^{\prime}(1-T+TH_{1}%
^{in}(z))\text{.}%
\]
Since there is a unique pgf that solves this equation, $H_{\ast}^{in}%
(z)=H_{1}(z)$. \ Thus, the in-component size distribution at the beginning of
an edge is the same for directed and undirected edges, and it is identical to
the distribution of component sizes at the end of an occupied edge in the bond
percolation model. \ 
\end{proof}

\begin{theorem}
$H^{in}(z)=H_{0}(z)$. \ 
\end{theorem}

\begin{proof}
Let $H_{\ast}^{in}(z)=H_{r}^{in}(z)=H_{u}^{in}(z)$. \ From equation
(\ref{Hin}), the probability generating function for the distribution of
in-component sizes in the percolation network is
\begin{align*}
H^{in}(z)  & =zG(H_{\ast}^{in}(z),1,H_{\ast}^{in}(z))\\
& =z\int_{0}^{\infty}\mathcal{G}(g(H_{\ast}^{in}(z),1,H_{\ast}^{in}%
(z)|\tau_{i}))dF(\tau_{i})\\
& =z\mathcal{G}(1-T+TH_{\ast}^{in}(z)).
\end{align*}
When $H_{1}(z)$ is substituted for $H_{\ast}^{in}(z)$ (which is justified by
the previous Lemma), this is identical to equation (\ref{H0}) for $H_{0}(z)$
in the bond percolation model. \ Since there is a unique pgf solution to this
equation, $H^{in}(z)=$ $H_{0}(z)$, so the distribution of in-components in the
percolation network is identical to the distribution of component sizes in the
bond percolation model.
\end{proof}

Since the mean size of out-components is equal to the mean size of
in-components in any semi-directed network, the bond percolation model
correctly predicts the mean size of outbreaks below the epidemic threshold.
\ Since the mean sizes of in- and out-components diverge simultaneously, the
bond percolation model also correctly predicts the critical transmissibility
$T_{c}$. \ Since the probability of having a finite in-component in the
percolation model is equal to the probability of being in a finite component
of the bond percolation model, the bond percolation model also correctly
predicts the final size of an epidemic. \ 

\section{Out-components}

In this section, we prove that the distribution of out-component sizes in the
epidemic percolation network for the SIR model from \cite{Newman1} is always
\textit{different} than the distribution of in-component sizes when there is a
nondegenerate distribution of infectious periods. \ As a corollary, we find
that the probability of an epidemic in the SIR model from the Introduction is
always less than or equal to its final size, with equality only when epidemics
have probability zero or the infectious period is constant. \ This is similar
to a result obtained by Kuulasmaa and Zachary \cite{Kuulasmaa2}, who found
that an SIR model defined on the $d$-dimensional integer lattice reduced to a
bond percolation process if and only if the infectious period is constant. \ 

The probability generating function for the total number of outgoing and
undirected edges incident to a node $i$ with infectious period $\tau_{i}$ is
\[
G(1,y,y|\tau_{i})=\mathcal{G}(g(1,y,y|\tau_{i}))=\mathcal{G}(1-T_{\tau_{i}%
}+T_{\tau_{i}}y),
\]
where $T_{\tau_{i}}$ is the conditional probability of transmission across
each edge given $\tau_{i}$, as defined in equation (\ref{Ti}). \ The number of
nodes we can reach by going forwards along edges starting from $i$ has a
Binomial($n_{i},T_{\tau_{i}}$) distribution. \ If we reach a node $j$ by
following an edge, then the number of nodes we can reach from $j$ by
continuing to go forwards (excluding the node from which we arrived) has a
binomial($k_{j}-1,T_{\tau_{j}}$) distribution. \ Unless $\tau_{i}$ is
constant, the out-components of the epidemic percolation network are not like
the components of a bond percolation model.

Suppose $i$ and $j$ are connected in the contact network. \ The conditional
transmission probability from $j$ to $i$ given $\tau_{i}$ is always $T$.
\ Thus, an edge across which we leave any node is directed (i.e., outgoing)
with probability $1-T$ and undirected with probability $T$. \ This allows us
to calculate the pgfs of the out-component distributions without
differentiating between outgoing and undirected edges:\ Let
\begin{align*}
G_{o}(x,y,u)  & =(1-T)G_{f}(x,y,u)+TG_{u}(x,y,u)\\
& =\frac{1}{\mathcal{G}^{\prime}(1)}\int_{0}^{\infty}\mathcal{G}^{\prime
}(g(x,y,u|\tau_{i}))dF(\tau_{i})
\end{align*}
be the probability generating function for the degree distribution of a node
that we reach by going forward along an outgoing or undirected edge (excluding
the edge along which we arrived). \ Let
\[
H_{\ast}^{out}(z)=(1-T)H_{f}^{out}(z)+TH_{u}^{out}(z)
\]
be the probability generating function for the size of the out-component at
the end of an outgoing or undirected edge. \ 

\begin{lemma}
For the SIR model from \cite{Newman1},
\begin{align*}
H_{f}^{out}(z)  & =zG_{f}(1,H_{\ast}^{out}(z),H_{\ast}^{out}(z)),\\
H_{u}^{out}(z)  & =zG_{u}(1,H_{\ast}^{out}(z),H_{\ast}^{out}(z)),\\
H^{out}(z)  & =zG(1,H_{\ast}^{out}(z),H_{\ast}^{out}(z)),
\end{align*}
and we have the following self-similarity equation:
\[
H_{\ast}^{out}(z)=zG_{o}(1,H_{\ast}^{out}(z),H_{\ast}^{out}(z)).
\]

\end{lemma}

\begin{proof}
From equation (\ref{gxyu|tau}), we have
\begin{align*}
g(1,(1-T)y+Tu,(1-T)y+Tu|\tau_{i})  & =1-T_{\tau_{i}}+T_{\tau_{i}%
}[(1-T)y+Tu]\\
& =g(1,y,u|\tau_{i})
\end{align*}
for all $y$, $u$, and $\tau_{i}$. \ This allows us to rewrite equation
(\ref{Hf,out}):
\begin{align*}
H_{f}^{out}(z)  & =zG_{f}(1,H_{f}^{out}(z),H_{u}^{out}(z))\\
& =\frac{z}{(1-T)\mathcal{G}^{\prime}(1)}\int_{0}^{\infty}\mathcal{G}^{\prime
}(g(1,H_{f}^{out}(z),H_{u}^{out}(z)|\tau_{i}))(1-T_{\tau_{i}})dF(\tau_{i})\\
& =\frac{z}{(1-T)\mathcal{G}^{\prime}(1)}\int_{0}^{\infty}\mathcal{G}^{\prime
}(g(1,H_{\ast}^{out}(z),H_{\ast}^{out}(z)|\tau_{i}))(1-T_{\tau_{i}}%
)dF(\tau_{i})\\
& =zG_{f}(1,H_{\ast}^{out}(z),H_{\ast}^{out}(z)).
\end{align*}
Similarly, we can rewrite equation (\ref{Hu,out}):
\begin{align*}
H_{u}^{out}(z)  & =zG_{u}(1,H_{f}^{out}(z),H_{u}^{out}(z))\\
& =\frac{z}{T\mathcal{G}^{\prime}(1)}\int_{0}^{\infty}\mathcal{G}^{\prime
}(g(1,H_{f}^{out}(z),H_{u}^{out}(z)|\tau_{i}))T_{\tau_{i}}dF(\tau_{i})\\
& =\frac{z}{T\mathcal{G}^{\prime}(1)}\int_{0}^{\infty}\mathcal{G}^{\prime
}(g(1,H_{\ast}^{out}(z),H_{\ast}^{out}(z)|\tau_{i}))T_{\tau_{i}}dF(\tau
_{i})\\
& =zG_{u}(1,H_{\ast}^{out}(z),H_{\ast}^{out}(z)).
\end{align*}
Finally, we can rewrite equation (\ref{Hout}):%
\begin{align*}
H^{out}(z)  & =zG(1,H_{f}^{out}(z),H_{u}^{out}(z))\\
& =z\int_{0}^{\infty}\mathcal{G}(g(1,H_{f}^{out}(z),H_{u}^{out}(z)|\tau
_{i}))dF(\tau_{i})\\
& =z\int_{0}^{\infty}\mathcal{G}(g(1,H_{\ast}^{out}(z),H_{\ast}^{out}%
(z)|\tau_{i}))dF(\tau_{i})\\
& =zG(1,H_{\ast}^{out}(z),H_{\ast}^{out}(z));
\end{align*}
but then
\begin{align*}
H_{\ast}^{out}(z)  & =(1-T)H_{f}^{out}(z)+H_{u}^{out}(z)\\
& =z[(1-T)G_{f}(1,H_{\ast}^{out}(z),H_{\ast}^{out}(z))+TG_{u}(1,H_{\ast}%
^{out}(z),H_{\ast}^{out}(z))]\\
& =zG_{o}(1,H_{\ast}^{out}(z),H_{\ast}^{out}(z)).
\end{align*}
As a corollary, we find that the analysis in Ref. \cite{Newman1} can be
corrected if we let $G_{0}(x)=G(1,x,x)$ and $G_{1}(x)=G_{o}(1,x,x)$ (see
equations (13) and (14) in \cite{Newman1}). \ 
\end{proof}

\begin{lemma}
$H_{\ast}^{in}(z)\leq H_{\ast}^{out}(z)$ for all $z\in\lbrack0,1]$. \ 
\end{lemma}

\begin{proof}
Since $\mathcal{G}^{\prime}$ is convex,
\begin{align*}
H_{\ast}^{out}(z)  & =zG_{o}(1,H_{\ast}^{out}(z),H_{\ast}^{out}(z))\\
& =\frac{z}{\mathcal{G}^{\prime}(1)}\int_{0}^{\infty}\mathcal{G}^{\prime
}(1-T_{\tau_{i}}+T_{\tau_{i}}H_{\ast}^{out}(z))dF(\tau_{i})\\
& \geq\frac{z}{\mathcal{G}^{\prime}(1)}\mathcal{G}^{\prime}(1-T+TH_{\ast
}^{out}(z))
\end{align*}
by Jensen's inequality. \ Equality holds only if $z=0$, $H_{\ast}^{out}(z)=1$,
$\mathcal{G}^{\prime}$ is constant, or $\tau_{i}$ is constant. \ Since
$H_{\ast}^{in}(z)$ is the solution to
\[
H_{\ast}^{in}(z)=\frac{z}{\mathcal{G}^{\prime}(1)}\mathcal{G}^{\prime
}(1-T+TH_{\ast}^{in}(z))\text{,}%
\]
we must have $H_{\ast}^{out}(z)\geq H_{\ast}^{in}(z)$. \ This can be seen by
fixing $z$ and considering the graphs of $y=zG_{o}(1,x,x)$ and $y=\frac
{z}{\mathcal{G}^{\prime}(1)}\mathcal{G}^{\prime}(1-T+Tx)$. \ $H_{\ast}%
^{out}(z)$ is the value of $x$ at which $y=zG_{o}(1,x,x)$ intersects the line
$y=x$. \ $H_{\ast}^{in}(z)$ is the value of $x$ at which $y=\frac
{z}{\mathcal{G}^{\prime}(1)}\mathcal{G}^{\prime}(1-T+Tx)$ intersects the line
$y=x$. \ Since $zG_{o}(1,x,x)\geq\frac{z}{\mathcal{G}^{\prime}(1)}%
\mathcal{G}^{\prime}(1-T+Tx)$, we must have $H_{\ast}^{out}(z)\geq H_{\ast
}^{in}(z)$. \ 
\end{proof}

\begin{theorem}
$H^{in}(z)\leq H^{out}(z)$ for all $z\in\lbrack0,1]$. \ Equality holds only
when $z=0$, $z=1$ and the percolation network is subcritical, or the
infectious period is constant.
\end{theorem}

\begin{proof}
From equation (\ref{Hin}),
\begin{align*}
H^{in}(z)  & =zG(H_{\ast}^{in}(z),1,H_{\ast}^{in}(z))\\
& =z\mathcal{G}(1-T+TH_{\ast}^{in}(z))\text{.}%
\end{align*}
From equation (\ref{Hout}),
\begin{align*}
H^{out}(z)  & =zG(1,H_{\ast}^{out}(z),H_{\ast}^{out}(z))\\
& =z\int_{0}^{\infty}\mathcal{G}(1-T_{\tau_{i}}+T_{\tau_{i}}H_{\ast}%
^{out}(z))dF(\tau_{i})\\
& \geq z\mathcal{G}(1-T+TH_{\ast}^{out}(z))\\
& \geq z\mathcal{G}(1-T+TH_{\ast}^{in}(z))\text{.}%
\end{align*}
The first inequality follows from the convexity of $\mathcal{G}$ and Jensen's
inequality. \ The second follows from the fact that $\mathcal{G}$ is
nondecreasing and $H_{\ast}^{out}(z)\geq H_{\ast}^{in}(z)$. \ Equality holds
in both inequalities only if $z=0$, $\mathcal{G}$ is constant, $H_{\ast}%
^{in}(z)=1$, or $\tau_{i}$ is constant. \ 
\end{proof}

Since the probability of an epidemic is $1-H^{out}(1)$ and the final size of
an epidemic is $1-H^{in}(1)$, it follows that the probability of an epidemic
is always less than or equal to its final size in the SIR model from
\cite{Newman1}. \ When the infectious period is constant, $H^{out}%
(z)=H^{in}(z)$ for all $z\in\lbrack0,1]$, so the in- and out-component size
distributions are identical and the probability and final size of an epidemic
are equal. \ When the infectious period has a nondegenerate distribution and
the percolation network is subcritical, $H^{out}(z)>H^{in}(z)$ for all
$z\in(0,1)$ (so the in- and out-components have dissimilar size distributions)
but $H^{out}(1)=H^{in}(1)=1$ (so the probability and final size of an epidemic
are both zero). \ If the network is supercritical and the infectious period is
nonconstant, $H^{out}(z)>H^{in}(z)$ for all $z\in\lbrack0,1]$, so in- and
out-components have dissimilar size distributions and the probability of an
epidemic is strictly less than its final size. \ 

Since the bond percolation model predicts the distribution of in-component
sizes, it cannot predict the distribution of out-component sizes or the
probability of an epidemic for any SIR model with a nonconstant infectious
period. \ However, it does establish an upper limit for the probability of an
epidemic in an SIR model. \ We have recently become aware of independent work
\cite{Miller} that shows similar results for more general sources of variation
in infectiousness and susceptibility in a model where these are independent
and uses Jensen's inequality to establish a lower bound for the probability
and final size of an epidemic. \ The lower bound corresponds to a site
percolation model with site occupation probability $T$, which is the model
that minimized the probability of no transmission in the Introduction. \ 

\section{Simulations}

In a series of simulations, the bond percolation model correctly predicted the
mean outbreak size (below the epidemic threshold), the epidemic threshold, and
the final size of an epidemic \cite{Newman1}. \ In Section 4, we showed that
the epidemic percolation network generates the same predictions for these quantities.

In Newman's simulations, the contact network had a power-law degree
distribution with an exponential cutoff around degree $\kappa$, so the
probability that a node has degree $k$ is proportional to $k^{-\alpha
}e^{-1/\kappa}$ for all $k\geq1$. \ This distribution was chosen to reflect
degree distributions observed in real-world networks
\cite{Newman1,Albert,Newman2,Newman3}. \ The probability generating function
for this degree distribution is
\[
\mathcal{G}(z)=\frac{\operatorname{Li}_{\alpha}(ze^{-1/\kappa})}%
{\operatorname{Li}_{\alpha}(e^{-1/\kappa})},
\]
where $\operatorname{Li}_{\alpha}(z)$ is the $\alpha$-polylogarithm of $z$.
\ In \cite{Newman1}, Newman used $\alpha=2$.

\begin{figure}
[ptb]
\begin{center}
\includegraphics[
height=4.7288in,
width=5.5486in
]%
{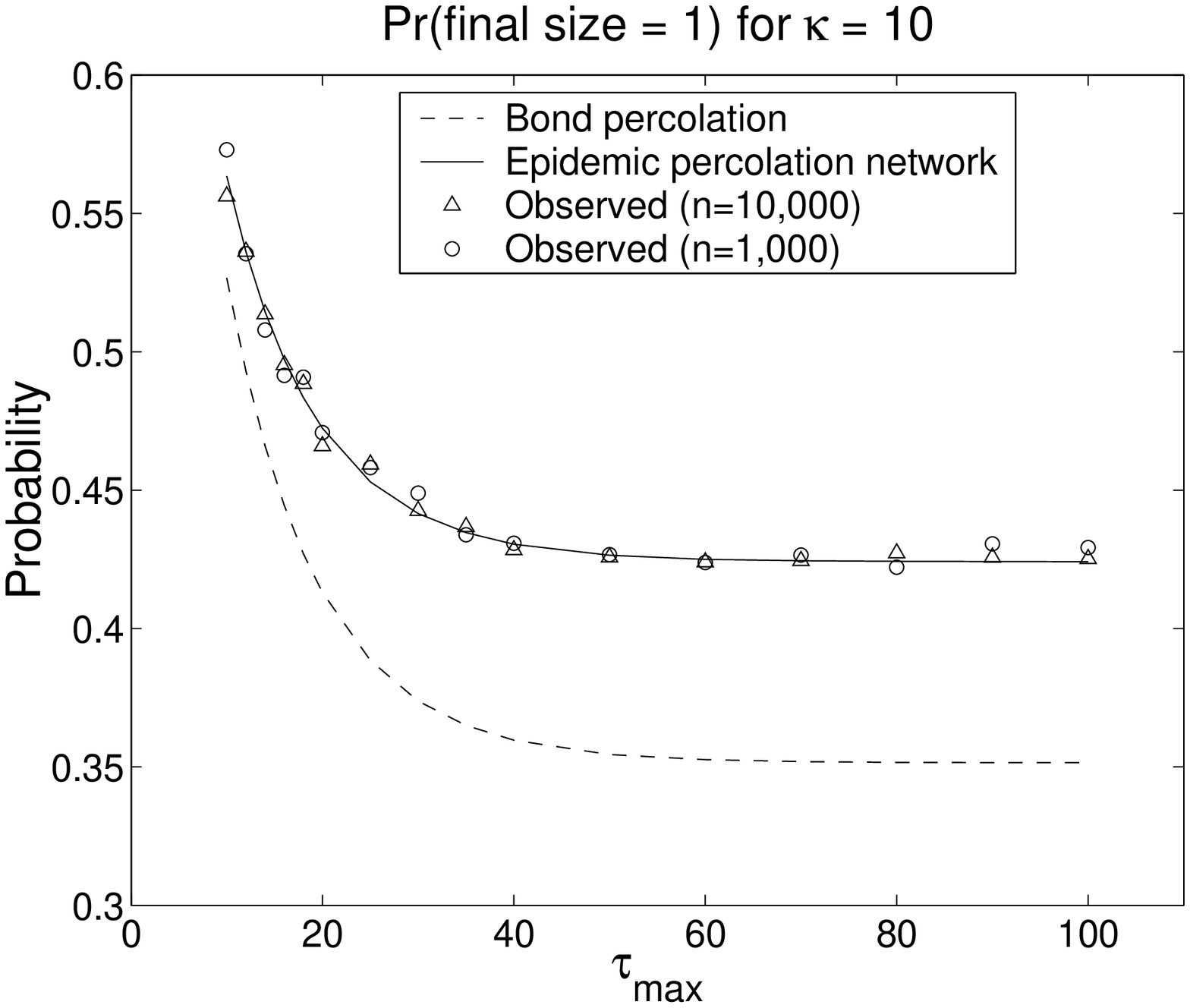}%
\caption{The predicted and observed probabilities of an outbreak of size one
on a contact network with $\kappa=10$ as a function of $\tau_{\max}$. \ Models
were run for $\tau_{\max}=10$, $12$, $14$, $16$, $18$, $20$, $25$, $30$, $35$,
$40$, $50$, $60$, $70$, $80$, $90$, $100$. \ Each observed value is based on
$10,000$ simulations in a population of size $n$. \ For $n=10,000$, $1,000$
simulations were conducted on each of $10$ contact networks. \ For $n=1,000$,
$100$ simulations were conducted on each of $100$ contact networks. \ }%
\label{k10p1}%
\end{center}
\end{figure}

\begin{figure}
[ptb]
\begin{center}
\includegraphics[
height=4.7288in,
width=5.5486in
]%
{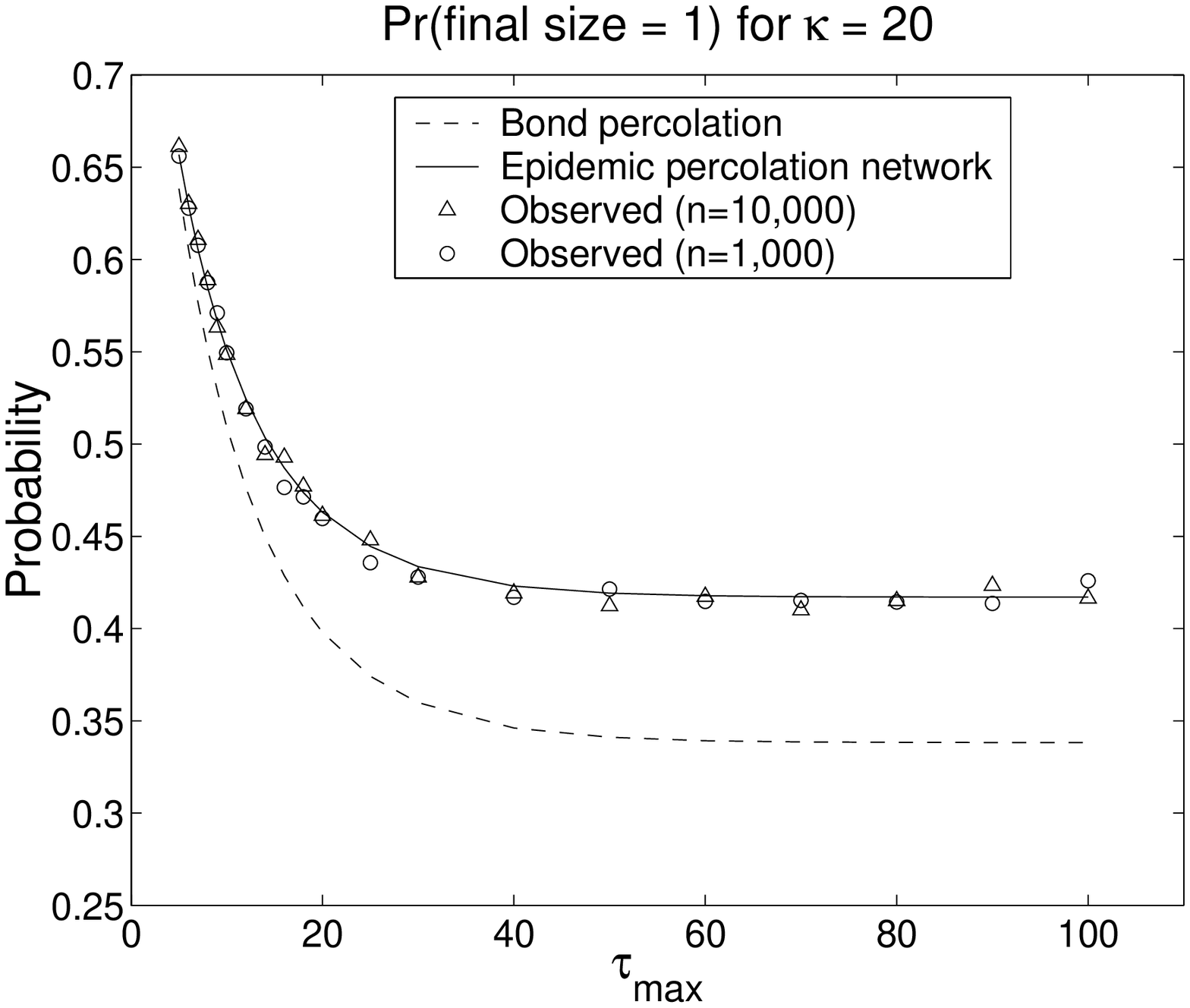}%
\caption{The predicted and observed probabilities of an outbreak of size one
on a contact network with $\kappa=20$ as a function of $\tau_{\max}$. \ Models
were run for $\tau_{\max}=5$, $6$, $7$, $8$, $9$, $10$, $12$, $14$, $16$,
$18$, $20$, $25$, $30$, $40$, $50$, $60$, $70$, $80$, $90$, and $100$. \ Each
observed value is based on $10,000$ simulations in a population of size $n$.
\ For $n=10,000$, $1000$ simulations were conducted on each of ten contact
networks. \ For $n=1000$, $100$ simulations were conducted on each of $100$
contact networks. \ }%
\label{k20p1}%
\end{center}
\end{figure}

\begin{figure}
[ptb]
\begin{center}
\includegraphics[
height=4.7288in,
width=5.5486in
]%
{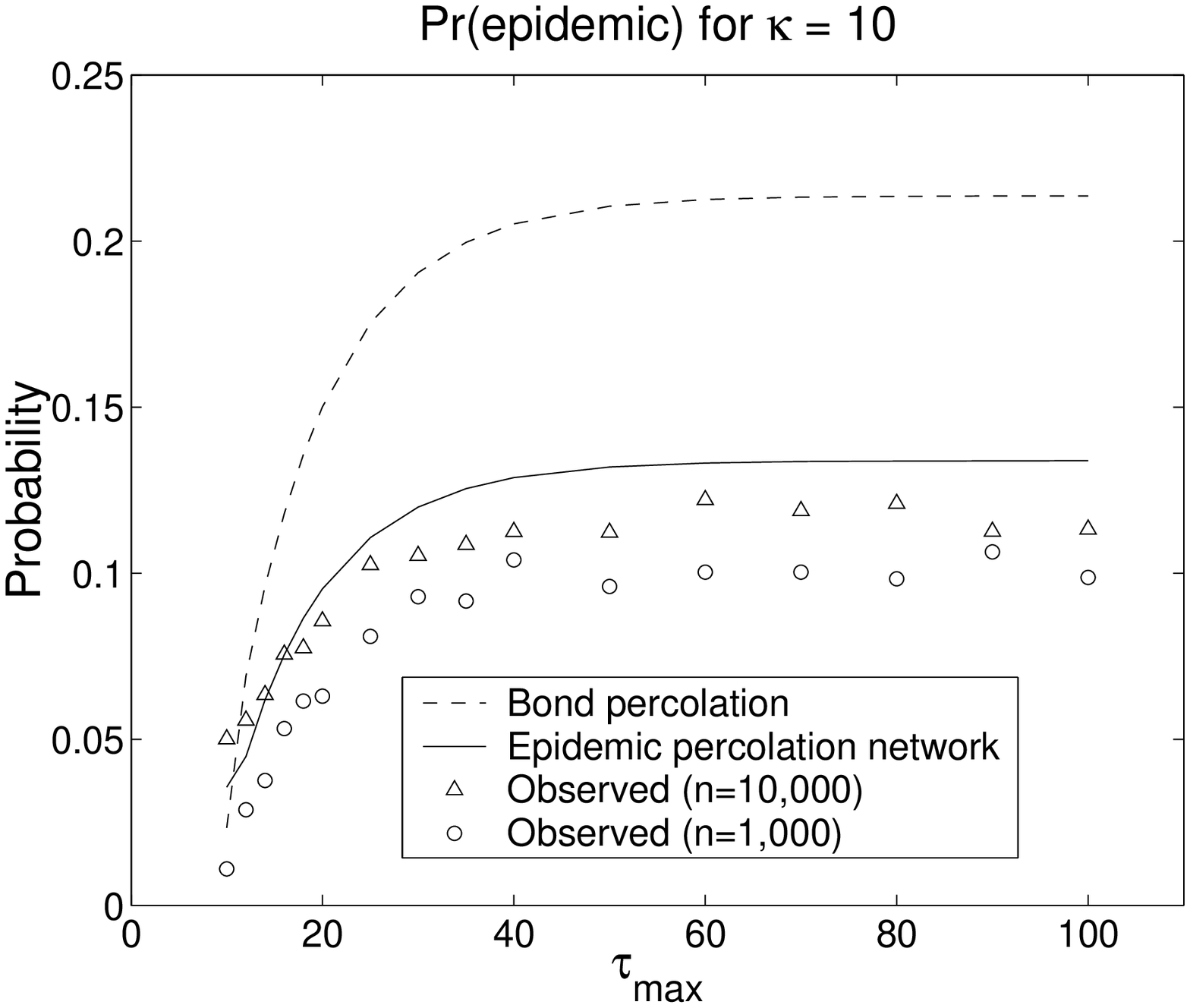}%
\caption{The predicted and observed probabilities of an epidemic on a contact
network with $\kappa=10$ as a function of $\tau_{\max}$. \ Models were run for
$\tau_{\max}=10$, $12$, $14$, $16$, $18$, $20$, $25$, $30$, $35$, $40$, $50$,
$60$, $70$, $80$, $90$, and $100$. \ Each observed value is based on $10,000$
simulations in a population of size $n$. \ For $n=10,000$, $1000$ simulations
were conducted on each of ten contact networks. \ For $n=1000$, $100$
simulations were conducted on each of $100$ contact networks. \ }%
\label{k10pepi}%
\end{center}
\end{figure}

\begin{figure}
[ptb]
\begin{center}
\includegraphics[
height=4.7288in,
width=5.5486in
]%
{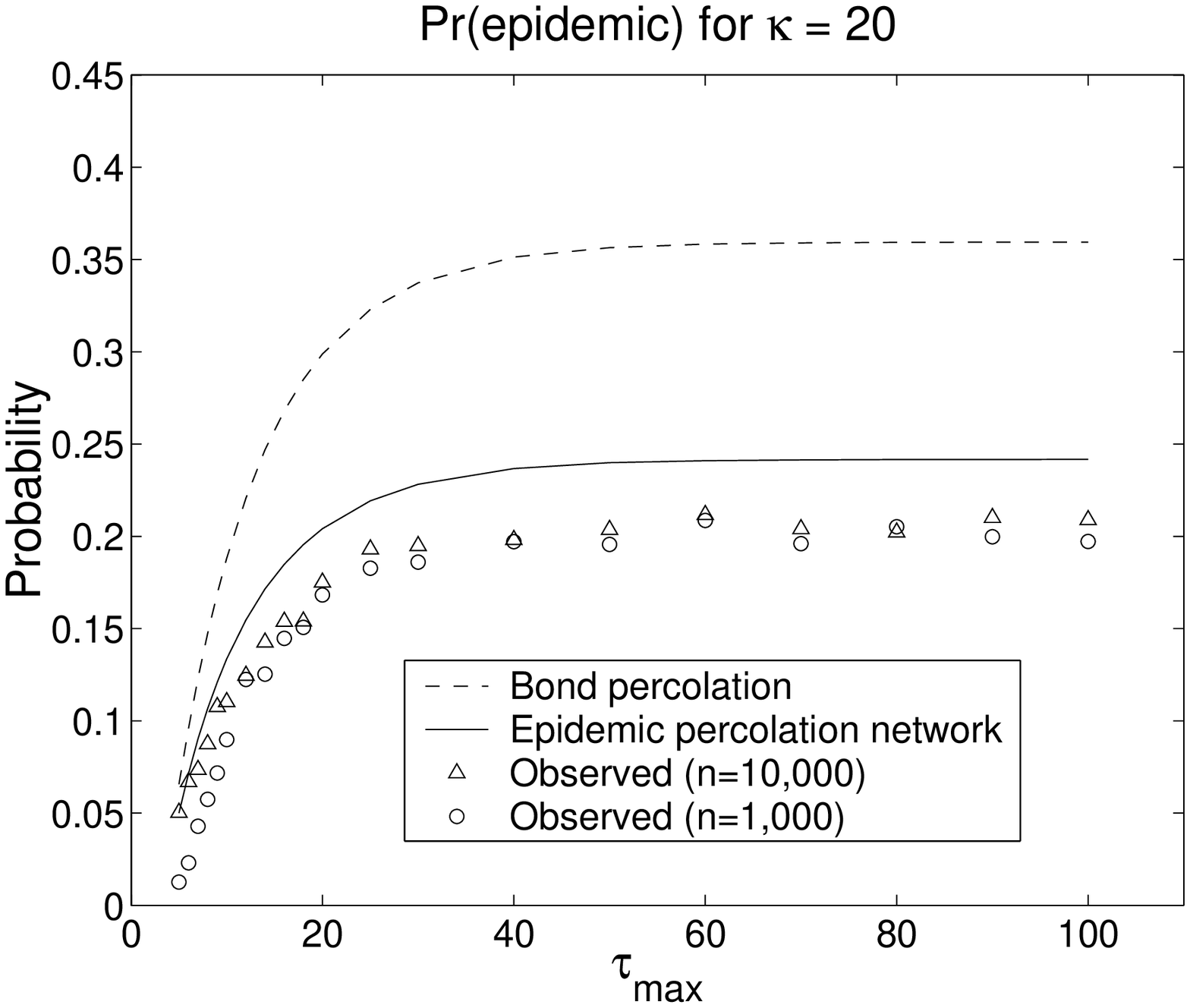}%
\caption{The predicted and observed probabilities of an epidemic on a contact
network with $\kappa=20$ as a function of $\tau_{\max}$. \ Models were run for
$\tau_{\max}=5$, $6$, $7$, $8$, $9$, $10$, $12$, $14$, $16$, $18$, $20$, $25$,
$30$, $40$, $50$, $60$, $70$, $80$, $90$, and $100$.\ \ Each observed value is
based on $10,000$ simulations in a population of size $n$. \ For $n=10,000$,
$1000$ simulations were conducted on each of ten contact networks. \ For
$n=1000$, $100$ simulations were conducted on each of $100$ contact networks.
}%
\label{k20pepi}%
\end{center}
\end{figure}

In our simulations, we retained the same contact network but used a contact
model adapted from the counterexample in the Introduction. \ We fixed
$\beta_{ij}=\beta_{0}=0.1$ for all $ij$ and let $\tau_{i}=1$ with probability
$0.5$ and $\tau_{i}=\tau_{\max}>1$ with probability $0.5$ for all $i$. \ The
predicted probability of an outbreak of size one is $G(1,0,0)$ in the epidemic
percolation network and $G(0,1,0)$ in the bond percolation model. \ The
predicted probability of an epidemic is $1-H^{out}(1)$ in the epidemic
percolation network and $1-H^{in}(1)$ in the bond percolation model. \ In all
simulations, an epidemic was declared when at least $100$ persons were
infected (this low cutoff produces a slight overestimate of the probability of
an epidemic in the simulations, favoring the bond percolation model).
\ Figures \ref{k10p1} and \ref{k20p1} show that percolation networks
accurately predicted the probability of an outbreak of size one for all
$(n,\kappa,\tau_{\max})$ combinations, whereas the bond percolation model
consistently underestimated these probabilities. \ Figures \ref{k10pepi} and
\ref{k20pepi} show that the bond percolation model significantly overestimated
the probability of an epidemic for all $(n,\kappa,\tau_{\max})$ combinations.
\ The percolation network predictions were far closer to the observed values. \ 

\section{Discussion}

For any time-homogeneous SIR epidemic model, the problem of analyzing its
final outcomes can be reduced to the problem of analyzing the components of an
epidemic percolation network. \ The distribution of outbreak sizes starting
from a node $i$ is identical to the distribution of its out-component sizes in
the probability space of percolation networks. \ Calculating this distribution
may be extremely difficult for a finite population, but it simplifies
enormously in the limit of a large population for many SIR\ models. \ For a
single randomly chosen imported infection in the limit of a large population,
the distribution of self-limited outbreak sizes is equal to the distribution
of small out-component sizes and the probability of an epidemic is equal to
the relative size of the GIN. \ For any finite set of imported infections, the
relative final size of an epidemic is equal to the relative size of the GOUT. \ 

In this paper, we used epidemic percolation networks to reanalyze the SIR
epidemic model studied in \cite{Newman1}. \ The mapping to a bond percolation
model correctly predicts the distribution of in-component sizes, the critical
transmissibility, and the final size of an epidemic. \ However, it fails to
predict the correct distribution of outbreak sizes and overestimates the
probability of an epidemic when the infectious period is nonconstant. \ Since
all known infectious diseases have nonconstant infectious periods and
heterogeneity in infectiousness has important consequences in real epidemics
\cite{Riley,Lipsitch,Dye}, it is important to be able to analyze such models correctly.\ \ 

The exact finite-population isomorphism between a time-homogeneous SIR model
and our semi-directed epidemic percolation network is not only useful because
it provides a rigorous foundation for the application of percolation methods
to a large class of SIR epidemic models (including fully-mixed models as well
as network-based models), but also because it provides further insight into
the epidemic model. \ For example, we used the mapping to an epidemic
percolation network to show that the distribution of in- and out-component
sizes in the SIR model from \cite{Newman1} could be calculated by treating the
incoming and outgoing infectious contact processes as separate directed
percolation processes, as in \cite{Miller}. \ However, in contrast with
\cite{Miller}, the semi-directed epidemic percolation network isolates the
fundamental role of the GSCC in the emergence of epidemics. \ The design of
interventions to reduce the probability and final size of an epidemic is a
central concern of infectious disease epidemiology. \ In a forthcoming paper,
we analyze both fully-mixed and network-based SIR models in which vaccinating
those nodes most likely to be in the GSCC is shown to be the most effective
strategy for reducing both the probability and final size of an epidemic. \ If
the incoming and outgoing contact processes are treated separately, the notion
of the GSCC is lost. \ 

\textbf{Acknowledgments: }\textit{This work was supported by the US National
Institutes of Health cooperative agreement 5U01GM076497 \textquotedblleft
Models of Infectious Disease Agent Study" (E.K.) and Ruth L. Kirchstein
National Research Service Award 5T32AI007535 \textquotedblleft Epidemiology of
Infectious Diseases and Biodefense" (E.K.), as well as a research grant from
the Institute for Quantitative Social Sciences at Harvard University (E.K.).
\ Joel C. Miller's comments on the proofs in Sections 3 and 4 were extremely
valuable, and we are also grateful for the comments of Marc Lipsitch, James H.
Maguire, and the anonymous referees of PRE. \ E.K. would also like to thank
Charles Larson and Stephen P. Luby of the Health Systems and Infectious
Diseases Division at ICDDR,B (Dhaka, Bangladesh).}

\appendix{}

\section{Epidemic percolation networks}

It is possible to define epidemic percolation networks for a much larger class
of stochastic SIR epidemic models than the one from \cite{Newman1}. \ First,
we specify an SIR model using probability distributions for recovery periods
in individuals and times from infection to infectious contact in ordered pairs
of individuals. \ Second, we outline time-homogeneity assumptions under which
the epidemic percolation network is well-defined. \ Finally, we define
infection networks and use them to show that the final outcome of the SIR
model depends only on the set of imported infections and the epidemic
percolation network. \ \ \ \ \ 

\subsection{Model specification}

Suppose there is a closed population in which every susceptible person is
assigned an index $i\in\{1,...,n\}$. \ A susceptible person is infected upon
infectious contact, and infection leads to recovery with immunity or death.
\ Each person $i$ is infected at his or her \textit{infection time} $t_{i}$,
with $t_{i}=\infty$ if $i$ is never infected. \ Person $i$ is removed (i.e.,
recovers from infectiousness or dies) at time $t_{i}+r_{i}$, where the
\textit{recovery period} $r_{i}$ is a random sample from a probability
distribution $f_{i}(r)$. \ The recovery period $r_{i}$ may be the sum of a
\textit{latent period}, when $i$ is infected but not yet infectious, and an
\textit{infectious period}, when $i$ can transmit infection. \ We assume that
all infected persons have a finite recovery period. \ Let $S(t)=\{i:t_{i}>t\}$
be the set of susceptible individuals at time $t$. \ Let $t_{(1)}\leq
t_{(2)}\leq...\leq t_{(n)}$ be the order statistics of $t_{1},...,t_{n}$, and
let $i_{(k)}$ be the index of the $k^{\text{th}}$ person infected. \ 

When person $i$ is infected, he or she makes infectious contact with person
$j\neq i$ after an \textit{infectious contact interval} $\tau_{ij}$. \ Each
$\tau_{ij}$ is a random sample from a conditional probability density
$f_{ij}(\tau|r_{i})$. \ Let $\tau_{ij}=\infty$ if person $i$ never makes
infectious contact with person $j$, so $f_{ij}(\tau|r_{i})$ has a probability
mass concentrated at infinity. \ Person $i$ cannot transmit disease before
being infected or after recovering, so $f_{ij}(\tau|r_{i})=0$ for all $\tau<0$
and all $\tau\in\lbrack r_{i},\infty)$. \ \ The \textit{infectious contact
time} $t_{ij}=t_{i}+\tau_{ij}$ is the time at which person $i$ makes
infectious contact with person $j$. \ If person $j$ is susceptible at time
$t_{ij}$, then $i$ infects $j$ and $t_{j}=t_{ij}$. \ If $t_{ij}<\infty$, then
$t_{j}\leq t_{ij}$ because person $j$ avoids infection at $t_{ij}$ only if he
or she has already been infected. \ 

For each person $i$, let his or her \textit{importation time} $t_{0i}$ be the
first time at which he or she experiences infectious contact from outside the
population, with $t_{0i}=\infty$ if this never occurs. \ Let $F_{0}%
(\mathbf{t}_{0})$ be the cumulative distribution function of the importation
time vector $\mathbf{t}_{0}=(t_{01},t_{02},...,t_{0n})$.

\subsection{Epidemic algorithm}

First, an importation time vector $\mathbf{t}_{0}$ is chosen. \ The epidemic
begins with the introduction of infection at time $t_{(1)}=\min_{i}(t_{0i})$.
\ Person $i_{(1)}$ is assigned a recovery period $r_{i_{(1)}}$. \ Every person
$j\in S(t_{(1)})$ is assigned an infectious contact time $t_{i_{(1)}j}%
=t_{(1)}+\tau_{i_{(1)}j}$. \ We assume that there are no tied infectious
contact times less than infinity. \ The second infection occurs at
$t_{(2)}=\min_{j\in S(t_{(1)})}\min(t_{0j},t_{i_{(1)}j})$, which is the time
of the first infectious contact after person $i_{(1)}$ is infected. \ Person
$i_{(2)}$ is assigned a recovery period $r_{i_{(2)}}$. \ After the second
infection, each of the remaining susceptibles is assigned an infectious
contact time $t_{i_{(2)}j}=t_{(2)}+\tau_{i_{(2)}j}$. \ The third infection
occurs at $t_{(3)}=\min_{j\in S(t_{(2)})}\min(t_{0j},t_{i_{(1)}j},t_{i_{(2)}%
j})$, and so on.\ \ After $k$ infections, the next infection occurs at
$t_{(k+1)}=\min_{j\in S(t_{(k)})}\min(t_{0j},t_{i_{(1)}j},...,t_{i_{(k)}j})$.
\ The epidemic stops after $m$ infections if and only if $t_{(m+1)}=\infty$. \ 

\subsection{Time homogeneity assumptions}

In principle, the above epidemic algorithm could allow the infectious period
and outgoing infectious contact intervals for individual $i$ to depend on all
information about the epidemic available up to time $t_{i}$. \ In order to
generate an epidemic percolation network, we must ensure that the joint
distributions of recovery periods and infectious contact intervals are defined
\textit{a priori}. \ The following restrictions are sufficient: \ 

\begin{enumerate}
\item We assume that the distribution of the recovery period vector
$\mathbf{r}=(r_{1},r_{2},...,r_{n})$ does not depend on the importation time
vector $\mathbf{t}_{0}$, the contact interval matrix $\mathbf{\tau}=[\tau
_{ij}]$, or the history of the epidemic.

\item We assume that the distribution of the infectious contact interval
matrix $\mathbf{\tau}$ does not depend on $\mathbf{t}_{0}$ or the history of
the epidemic. \ 
\end{enumerate}

With these time-homogeneity assumptions, the cumulative distributions
functions $F(\mathbf{r})$ of recovery periods and $F(\mathbf{\tau}%
|\mathbf{r})$ of infectious contact intervals are completely specified
\textit{a priori}. \ Given$\ \mathbf{r}$ and $\mathbf{\tau}$, the epidemic
percolation network is a semi-directed network in which there is a directed
edge from $i$ to $j$ iff $\tau_{ij}<\infty$ and $\tau_{ji}=\infty$, a directed
edge from $j$ to $i$ iff $\tau_{ij}=\infty$ and $\tau_{ji}<\infty$, and an
undirected edge between $i$ and $j$ iff $\tau_{ij}<\infty$ and $\tau
_{ji}<\infty$. \ The entire time course of the epidemic is determined by
$\mathbf{r}$, $\mathbf{\tau}$, and $\mathbf{t}_{0}$. \ However, its final size
depends only on the set $\{i:t_{0i}<\infty\}$ of possible imported infections
and the epidemic percolation network corresponding to $\mathbf{\tau}$. \ In
order to prove this, we first define the \textit{infection network,} which
records the chain of infection from a single realization of the epidemic model.\ 

\subsection{Infection networks}

Let $v_{i}$ be the index of the person who infected person $i$, with $v_{i}=0$
for imported infections and $v_{i}=\infty$ for uninfected nodes. \ If tied
finite infectious contact times are possible, then choose $v_{i}$ from all $j$
such that $t_{ji}=t_{i}$. \ The infection network has the edge set
$\{v_{i}i:0<v_{i}<\infty\}$. \ It is a purely directed subgraph of the
epidemic percolation network corresponding to $\mathbf{\tau}$ because
$\tau_{v_{i}i}<\infty$ for every edge $v_{i}i$. \ Since each node has at most
one incoming edge, all components of the infection network are trees or
isolated nodes. \ Every imported case is either the root node of a tree or an
isolated node. \ Every person infected through transmission within the
population is a nonroot node in a tree. \ Uninfected persons are isolated nodes.

The infection network can be represented by a vector $\mathbf{v}%
=(v_{1},..,v_{n})$, as in Ref. \cite{Wallinga}. \ If $v_{j}=0$, then
$t_{j}=t_{0j}$. \ If $0<v_{j}<\infty$, then $j$ is in a component of the
infection network with a root node $imp_{j}$ and its infection time is
\[
t_{j}=t_{imp_{j}}+\sum_{k=1}^{m}\tau_{i_{k}j_{k}},
\]
where the edges $i_{1}j_{1},...,i_{m}j_{m}$ form a directed path from
$imp_{j}$ to $j$. \ This path is unique because all nontrivial components of
the infection network are trees. \ If $v_{j}=\infty$, then $t_{j}=\infty$.
\ The removal time of each node $i$ is $t_{i}+r_{i}$. \ If there is more than
one possible infection network, they must all be consistent with
$(t_{1},...,t_{n})$ by definition of $v_{i}$. \ Therefore, the entire time
course of the epidemic is determined by the importation time vector
$\mathbf{t}_{0}$, the recovery period vector $\mathbf{r}$, and the infectious
contact interval matrix $\mathbf{\tau}$. \ \ \ \ \ 

\subsection{Final outcomes and epidemic percolation networks}

\begin{theorem}
In an epidemic with infectious contact interval matrix $\mathbf{\tau}$, a node
is infected if and only if it is in the out-component of a node $i$ with
$t_{0i}<\infty$ in the percolation network. \ (Equivalently, a node is
infected if and only if its in-component includes a node $i$ with
$t_{0i}<\infty$.) \ Therefore, the final outcome of the SIR model depends only
on the set of imported infections and the epidemic percolation network
corresponding to $\mathbf{\tau}$. \ 
\end{theorem}

\begin{proof}
Suppose that person $j$ is in the out-component of a node $i$ with
$t_{0i}<\infty$ in the epidemic percolation network corresponding to
$\mathbf{\tau}$. \ Then there is a sequence $i_{1}j_{1},...,i_{m}j_{m}$ such
that $i_{1}=i$, $j_{m}=j$, and $\tau_{i_{k}j_{k}}<\infty$ for $1\leq k\leq m$,
so
\[
t_{j}\leq t_{0i}+\sum_{k=1}^{m}\tau_{i_{k}j_{k}}<\infty,
\]
and $j$ must be infected during the epidemic. \ Now suppose that $t_{j}%
<\infty$. \ Then there exists an imported case $i$ and a sequence $i_{1}%
j_{1},...,i_{m}j_{m}$ such that $i_{1}=i$, $j_{m}=j$, and
\[
t_{j}=t_{i}+\sum_{k=1}^{m}\tau_{i_{k}j_{k}}.
\]
Since $t_{j}<\infty$, it follows that $\tau_{i_{k}j_{k}}<\infty$ for all $k$.
\ But then the epidemic percolation network corresponding to $\mathbf{\tau}$
has an edge with the proper direction or an undirected edge between $i_{k}$
and $j_{k}$ for all $k$, so $j$ is in the out-component of $i$. \ \ 
\end{proof}

By the law of iterated expectation (conditioning on $\mathbf{\tau})$, this
result implies that the distribution of outbreak sizes caused by the
introduction of infection to node $i$ is identical to the distribution of his
or her out-component sizes in the probability space of epidemic percolation
networks. \ Furthermore, the probability that person $i$ gets infected in an
epidemic is equal to the probability that his or her in-component contains at
least one imported infection. \ This isomorphism holds in any finite
population. \ In the limit of a large population, the probability that node
$i$ is infected in an epidemic is equal to the probability that he or she is
in the GOUT and the probability that an epidemic results from the infection of
node $i$ is equal to the probability that he or she is in the GIN. \ This
logic can be extended to predict the mean size of self-limited outbreaks and
the probability and final size of an epidemic for outbreaks started by any
given set of imported infections. \ 

\section{Figures and tables}%

%

%

%

%

\end{document}